\documentstyle[12pt]{article}
\begin{document}
\def \d {{\rm d}}

\title{
 Smearing of chaos in sandwich {\em pp\,}-waves}
\author{
J. Podolsk\'y\thanks{E--mail: {\tt podolsky@mbox.troja.mff.cuni.cz}}
,\  K. Vesel\'y
\\ \\ Department of Theoretical Physics,\\
Charles University, Faculty of Mathematics and Physics\\
V Hole\v{s}ovi\v{c}k\'ach 2, 180~00 Prague 8, Czech Republic.\\ }
\maketitle

\baselineskip=19pt

\begin{abstract}
Recent results demonstrating the chaotic behavior of
geodesics in non-homogeneous vacuum {\it pp\,}-wave solutions are
generalized. Here we concentrate on motion in non-homogeneous sandwich
{\it pp\,}-waves and show that chaos smears as the duration of
these gravitational  waves is reduced. As the number of radial bounces
of any geodesic decreases, the outcome channels to infinity become fuzzy,
and thus the fractal structure of the initial conditions characterizing
chaos is cut at lower and lower levels.  In the limit of
impulsive waves, the motion is fully non-chaotic. This is proved by
presenting the geodesics in a simple explicit form which permits a
physical interpretation, and demonstrates the focusing effect. It
is shown that a circle of test particles is deformed by the impulse into
a family of closed hypotrochoidal curves in the transversal plane. These are
deformed in the longitudinal direction in such a way that a specific
closed caustic surface is formed.
\end{abstract}

\bigskip

\newpage

\section{Introduction}

The widely known class of  plane-fronted gravitational waves with
parallel rays ({\it pp\,}-waves) has become a paradigm of exact
radiative spacetimes in general relativity. Found by Brinkmann
in 1923 \cite{Brink} and later rediscovered independently
by several authors, it has been studied
for decades. The metric of vacuum {\it pp\,}-waves
can be written in the standard form \cite{KSMH}
\begin{equation}
\d s^2=2\,\d\zeta \d\bar\zeta-2\,\d u\d v-(f+\bar f)\,\d u^2\ , \label{E1}
\end{equation}
where  $f(u, \zeta)$ is an arbitrary function of the retarded time
$u$ and the complex coordinate $\zeta$ spanning the plane wave surfaces
$u=u_0=$ const.
When $f$ is linear in $\zeta$, the metric (\ref{E1}) represents
Minkowski universe since the only non-trivial components of the
curvature tensor are proportional to $f_{,\zeta\zeta}$.
Thus the simplest case for which (\ref{E1}) describes
gravitational waves arises for $f=h(u)\zeta^2$,
where  an arbitrary function $h(u)$
characterizes the `profile' of the wave.
Solutions of this type are called plane waves (or homogeneous
{\it pp\,}-waves) and have been investigated
extensively in the literature (see \cite{KSMH},\cite{BGM} for Refs).
This important class of exact radiative spacetimes has also
been used for the construction of sandwich waves (see e.g.
\cite{BPR}-\cite{PV1}) for which $h(u)$ is non-vanishing
on a finite interval of $u$ only. Also, the very
interesting problem of the collision of two plane waves has been
thoroughly studied (see \cite{Grif} for review of the topic and
Refs.)

Surprisingly, more general non-homogeneous {\it pp\,}-waves with generic or
sandwich wave-profiles have not been investigated as much
(a physical interpretation of some solutions of this type was
proposed e.g. in \cite{Bon},\cite{Les}). On the other
hand, much attention has been payed to impulsive waves described by
the metric (\ref{E1}) with the function $f$ of the
form $f=\delta(u)F(\zeta)$, where $\delta(u)$ is the Dirac
distribution. These solutions have attracted attention
recently not only for the investigation of gravitational radiation
emitted during the collision of black holes \cite{DE78},\cite{DEP93},
but also in the context of field and string theories in a study of
the scattering processes at extremely high Planckian energies, see
\cite{tHooft} and elsewhere.
Such impulsive waves can naturally be understood as
distributional limits of  suitable sequences of sandwich waves
regularizing the Dirac $\delta$.
Spacetimes of this type can also be obtained by boosting
the Schwarzschild, other Kerr-Newman, and dilaton black holes
\cite{AiSe}-\cite{CJS},
or axially symmetric solutions of the Weyl class \cite{R4} to
the speed of light. Some of these solutions can be interpreted as
impulsive {\it pp\,}-waves generated by null particles of
an arbitrary multipole structure \cite{GrPo97}.
Another method for constructing general impulsive
{\it pp\,}-wave spacetimes has been proposed in \cite{Penrose}: the
`scissors-and-paste' approach is based on the removal of a null
hyperplane from Minkowski spacetime and re-attaching the parts by making
a formal identification with a `warp'. This corresponds to a
specific coordinate shift \cite{DH}.

Impulsive {\it pp\,}-waves can alternatively be introduced using
a different coordinate system in which the metric is explicitly
continuous \cite{DE78}, \cite{Penrose}-\cite{PoVe98}.
It has only recently been shown \cite{KuSt99} that these
approaches are equivalent in a rigorous sense. They lead to identical
(unique) particle motion, i.e. the corresponding geodesics in
(\ref{E1}), see \cite{Bal97}-\cite{KuSt98}, agree with those
obtained from the continuous form of the metric.

In our recent works \cite{PVcha1}, \cite{PVcha2} we demonstrated that geodesic
motion in the class of non-homogeneous vacuum {\it pp\,}-wave
spacetimes is chaotic. This is the first example of
chaos in exact radiative spacetimes (chaotic behavior of geodesics
in various black-hole spacetimes is well known, see e.g.
\cite{Hob}-\cite{CG}, and references therein).
It is the purpose of the presented paper to generalize these
results which remained restricted to spacetimes
(\ref{E1}) with $f=\frac{2}{n}C\zeta^n$, $n=3, 4, \cdots$, i.e. those
having a {\it constant} profile, $h(u)=C=$ const. In particular, we wish
to investigate the behavior of geodesics in non-homogeneous sandwich
{\it pp\,}-waves and in the corresponding impulsive limits.

In section~2 we briefly summarize previous results concerning chaotic
motion in {\it pp\,}-waves  with a constant profile. Also, we introduce
the concepts and quantities necessary for our subsequent investigation.
Section~3 is devoted to geodesics in the simplest sandwich {\it pp\,}-waves
having a `square' profile.
A rigorous characterization of the level of the fractal structure
of basin boundaries (which separate initial conditions leading to different
outcomes) by the `number of radial bounces' of
geodesics is introduced. This enables us to describe quantitatively how
chaos smears as the sequence of sandwich waves approaches the impulsive
limit. The same behavior of geodesics is investigated within the class of
smooth sandwich waves in section~4, where a relation between the smearing
of chaotic behavior and the gradual vanishing of the outcome channels
is also discussed. In section~5 explicit (i.e. non-chaotic) geodesics in the
resulting non-homogeneous impulsive {\it pp}-waves are presented and
interpreted physically. Finally, in section~6 we discuss the effect of
the focusing of geodesics and describe the deformation of a disc of test
particles.

\section{Chaotic motion in non-homogeneous {\it pp\,}-waves
with a constant profile}

As shown in \cite{PVcha1}, \cite{PVcha2}
the geodesic equations for (\ref{E1}) reduce to
\begin{eqnarray}
&&\ddot\zeta + \textstyle{\frac{1}{2}}\bar f_{,\bar\zeta}\, U^2 = 0 \ ,\label{E2}\\
&&u(\tau)=U\tau+\tilde U \ ,\label{E3}\\
&&v(\tau)=\textstyle{\frac{1}{2}}U^{-1}\int\,[\,2\dot\zeta \dot{\bar\zeta}
  -(f+\bar f)\,U^2-\epsilon\,]\,\d \tau+\tilde V \ ,\label{E4}
\end{eqnarray}
where $\tau$ is an affine parameter, dot denotes $d/d\tau$, $U, \tilde U, \tilde V$
are constants, and $\epsilon=-1, 0, +1$ for timelike, null or spacelike geodesics,
respectively. It suffices to find solutions of Eq. (\ref{E2}).
Introducing real coordinates $x$ and $y$
by $\zeta=x+\hbox{i}y$, this system follows from the Hamiltonian
\begin{equation}
H=\textstyle{\frac{1}{2}}\left(p_x^2+p_y^2 \right)+V(x, y, u)\ , \label{E7}
\end{equation}
where the potential is $V(x,y,u)=\textstyle{\frac{1}{2}}U^2\, {\cal R}e\,f$.
For non-homogeneous {\it pp\,}-waves given by
\begin{equation}
f=\textstyle{\frac{2}{n}}\,h(u)\,\zeta^n\ ,\quad n=3, 4, \cdots\ , \label{E7a}
\end{equation}
the corresponding polynomial potential
\begin{equation}
V(x, y, u) =\textstyle{\frac{1}{n}}U^2h(u)\, {\cal R}e\,\zeta^n \label{E7b}
\end{equation}
at any $u=u_0$ is called `$n$-saddle'. It can by visualized in polar
coordinates $\rho$, $\phi$ where $\zeta=\rho\exp(\hbox{i}\phi)$,
in which it takes the form
$V(\rho,\phi, u_0)=\frac{1}{n}U^2h(u_0)\, \rho^n\cos(n\phi)$.

It was shown in a series of mathematical papers
\cite{Rod}-\cite{CR1} that motion in the Hamiltonian system
(\ref{E7}) with a polynomial potential (\ref{E7b})
where $h(u)=C=$ const. is chaotic.
Note that, interestingly, in the simplest case $n=3$
the corresponding `monkey saddle' potential (after removing
the factor $CU^2$) is $V(x, y) = \frac{1}{3}x^3-xy^2$,
so that we get the particular case of the famous H\'enon-Heiles Hamiltonian
\cite{HH} (with missing quadratic terms) which is a `textbook'
example of a chaotic system. This was investigated by Rod \cite{Rod}
who concentrated on the topology of {\it bounded orbits} in the energy
manifolds $H=E>0$.
The sets of orbits asymptotic to the basic unstable periodic orbits
(denoted by $\Pi_j$) as $\tau\to\pm\infty$ intersect transversely.
This proves the existence of homoclinic and heteroclinic orbits
and indicates the complicated structure of the flow.
These results were later refined in \cite{RPC} by showing that
$\Pi_j$ are hyperbolic, so that they
admit stable and unstable asymptotic manifolds. Finally, in
\cite{CR3} the above Hamiltonian was presented
as an example of a system for which the Smale horseshoe map can
explicitly be embedded as a subsystem along the homoclinic
and heteroclinic orbits.
It was shown in \cite{Rod} - \cite{CR1} that similar results
hold for a general potential (\ref{E7b}) with
$h=C$. Therefore, geodesic motion in all non-homogeneous {\it
pp\,}-wave spacetimes with the corresponding function (\ref{E7a})
is chaotic.

In order to support these arguments for the chaotic behavior, we
investigated \cite{PVcha1}--\cite{PVcha2} the structure of motion by
a fractal method. Complementary to the analysis described above,
we concentrated on {\it unbounded} geodesics. The fractal method
(advanced in relativity in \cite{Conto1}-\cite{CG} and elsewhere)
starts with a definition of several
distinct asymptotic outcomes (given here by `types of ends' of all
trajectories). Subsequently, a set of initial conditions is
evolved until one of the outcome states is
reached. Chaos is established if the basin boundaries which separate
initial conditions leading to different outcomes are fractal.
Such fractal partitions are the result of chaotic dynamics and
measure an extremally sensitive dependence of the evolution on the
choice of initial conditions. We have observed exactly these structures
in the system studied.
We integrated numerically the equations of motion given by
(\ref{E2}), (\ref{E7a}) for $h=C$.
The initial conditions (without loss of generality) were chosen such
that the geodesics started (from rest) at $\tau=0$ from a unit circle in the
($x, y$)-plane. The initial positions were parametrized
by an angle $\phi\in [-\pi,\pi)$ such that $x(0)=\cos\phi$, $y(0)=\sin\phi$.
In Fig.~1 we present typical trajectories for $n=3, 4, 5$.
Each unbounded geodesic escapes to infinity
where the curvature singularity is located
{\it only along } one of the $n$ {\it distinct outcome channels}
in the potential with the radial axis $\phi_j=(2j-1)\pi/n$, $j=1, \cdots, n$
(in fact, it oscillates around the axis with frequency growing
to infinity and amplitude approaching zero \cite{PVcha1}). These
channels represent possible outcomes of our system and we label them
by the corresponding values of $j$. In certain regions the function
$j(\phi)$ representing a portrait of the basin structure depends
sensitively on initial position given by $\phi$ --- see Fig.~1 where the
(finite resolution) results for $n=3, 4, 5$ are shown.
In the same diagrams we plot the function $\tau_s(\phi)$ which takes
the value of $\tau$ when the singularity is reached by a given geodesic.
The boundaries between the outcomes are fractal
which we confirmed in \cite{PVcha2} on the enlarged detail, on the
detail of the detail etc. up to the sixth level
 (where numerical errors became significant).
At {\it each level} the structure has the property that
between two connected  sets of geodesics  with
channels $j_1$ and $j_2\not=j_1$ there is always
a smaller set of geodesics with channel  $j_3$ such
that $j_3\not=j_1$ and $j_3\not=j_2$.
This has a counterpart in $\tau_s(\phi)$, see Fig.~1.
The value of $\tau_s$ diverges on each discontinuity
of $j(\phi)$, i.e., on any fractal basin boundary.
There is an infinite number of peaks, each corresponding to an
unstable trapped orbit which never `decides' on a particular outcome
to infinity. Also, $\tau_s$ increases as one zooms into the higher
levels of the fractal. This is natural since higher
levels are generated by geodesics which undergo `more bounces'
in the inner region before escaping through
one of the outcome channels.

\section{Motion in shock waves and smearing of chaos in the impulsive
limit of simplest sandwich waves}

It can immediately be observed that the above results can easily
be applied to a description of geodesics in shock {\it pp\,}-waves
given by the metric (\ref{E1}), (\ref{E7a}) with
$h(u)=C\Theta(u)$, where $\Theta(u)$ is the
Heaviside step function. It is natural to consider free test
particles which are at rest ($\dot x=0=\dot y$) in the flat
Minkowski half-space $u<0$. (The Minkowski coordinates
are given by $x_M=\sqrt2\,x$, $y_M=\sqrt2\,y$, $z_M=(v-u)/\sqrt2$,
$t_M=(v+u)/\sqrt2$.) At $u=0$ these particles are hit by
the shock and subsequently for $u>0$ they move in
the wave with a constant profile $h=C$. The behavior of the corresponding
geodesics has been summarized in section~2. In particular, the trajectories
and the fractal structure of basin boundaries have again the form indicated
in Fig.~1. Therefore, the geodesic motion in non-homogeneous shock
{\it pp\,}-waves is chaotic in a rigorous sense.

However, it could be argued that these above results concern
very specific and rather `unrealistic' classes of
{\it pp\,}-wave solutions for which the profile function $h(u)$ in
(\ref{E7a}) is constant on an infinite interval of the
retarded time $u$. Such waves have an `infinite
duration' and a constant `strength'. It is the purpose of the
presented work to investigate more realistic non-homogeneous
{\it pp\,}-waves, namely sandwich waves of this type
described by functions $h(u)$ having only finite support.

For an investigation of sandwich waves it is convenient to
parametrize the geodesics by the coordinate $u$ instead of the
parameter $\tau$; for $\zeta(u)$ we get from Eqs. (\ref{E2}),
(\ref{E3}), (\ref{E7a})
\begin{equation}
\zeta'' + h(u)\bar\zeta^{n-1} = 0 \
,\label{E9a}
\end{equation}
where prime denotes $d/d u$. For timelike geodesics we can
simply substitute $u=U\tau+\tilde U$ in the result in order to
obtain the dependence on the proper time.

It is natural to start with the simplest sandwich waves having a
`square' profile
\begin{equation}
h(u) =\frac{1}{\tilde a}\Big[\Theta(u)-\Theta(u-a)\Big]\ , \label{E9}
\end{equation}
where $a$ and $\tilde a$ are positive constants. There are flat Minkowski
regions in front of the wave $(u<0)$ and behind the wave $(u>a)$. Within
the wavezone $(0<u<a)$ the amplitude is constant, $h=1/\tilde a$,
and we can use the results described above. In particular,
we can study a deformation of a ring of particles in the
$(x,y)$-plane which are at rest in front of the wave. The
particles start moving at $u=0$ and then follow exactly the same
trajectories $\zeta(u)$ as shown in Fig.~1 for $u<a$. At $u=a$ the influence
of the sandwich wave ends, the potential (\ref{E7b}) defining the
outcome channels vanishes, and for $u>a$, the particles move uniformly
in different directions along straight lines in the flat space behind
the wave. Consequently, the fractal structure of basin boundaries indicating
chaos is `cut' at some level given by the value of $a$: there is not enough
time for particles to bounce {\it arbitrarily many} times
between the potential walls. As $a\to0$, the number of bounces
tends to zero so that the fractal structure is completely erased.
In other words, for smaller $a$, the dependence on initial conditions is
less sensitive since the prediction of the outcome  can be done with
only a lower resolution. The narrower the sandwich wave, the `less
chaotic' the corresponding geodesic motion. Finally, in the
impulsive case given by the limit $a=\tilde a\to0$ of (\ref{E9}), i.e.
$f=\frac{2}{n}\delta(u)\zeta^n$, $n\ge3$, the motion is non-chaotic.
This effect can be called a `smearing of chaos' in the
impulsive limit of non-homogeneous sandwich {\it pp\,}-waves.

This behavior can be described formally. Let us define the {\it
number of bounces} $N$ of the geodesic as the number of (local)
maxima of the function  $\rho(u)=|\zeta(u)|=\sqrt{x^2(u)+y^2(u)}$
measuring the radial distance from the origin of the
($x,y$)-plane. $N$ represents the number of times the geodesic
crosses the phase-space surface of section $\rho'=0$ (with  $\rho''<0$)
before escaping to infinity. This is a more appropriate measure here
than the twist number used in literature on chaotic dynamics
\footnote{Note that actually the particle does
{\it not stop} at this section --- only its {\it radial} velocity
vanishes --- so that the event could better be called a `turn' or
a `radial deflection' rather then a `bounce' on the potential
wall, but we use this last word for its natural intuitive
meaning.}.
As in the previous section, we consider a family of
geodesics starting from rest from a unit circle $\rho(0)=1$ with
the angle $\phi$ parametrizing their initial positions. In Fig.~2
we plot the function $N(\phi)$ for the geodesics given by Eq.
(\ref{E9a}) with $h=1$ and $n=3$.  The sequence of
graphs shows the zooming in of the fractal interval
around the value $\phi\approx0$. Also, we plot the function $u_s(\phi)$
which takes the value of $u$ when the singularity at $\rho=\infty$
is reached by a given geodesic. It is obvious that the fractal
structure described by the function $N(\phi)$ corresponds
to the structure given by $u_s(\phi)$. Also, $N$ is a precise
definition of the `level' of the fractal since we have demonstrated
that the $N$-th level of the fractal structure is given by those geodesics
which  bounce exactly $N$ times before they choose one of the outcome
channels. Now we can describe how the fractal structure
indicating chaos arises. Let us denote by $I_k$ the set of initial conditions
$\phi$ generating geodesics with $N(\phi)\le k$,
where $k=0, 1, 2, \cdots$. Clearly, $I_0\subset I_1\subset I_2\subset
\cdots\subset \langle 0, 2\pi)$, and the complement of
$\lim_{k\to\infty} I_k$ describes the fractal boundary
intimately related to the fractal structure of the
outcome basin boundaries (although the discontinuities in
$N(\phi)$ and $j(\phi)$ do not coincide). Now, we define a sequence
of real numbers $a_k=\max\{u_s(\phi)\  |\  \phi\in I_k\}$. Their
physical meaning is the following: at $u>a_k$ {\it all}
geodesics generated by $I_k$ have already fallen into the
singularity at $\rho=\infty$ along the three channels
given by $\phi_j$ (after performing at most $k$ bounces).
From numerical simulations visualized in Fig.~2
we obtained the values $a_0=6.3$, $a_1=9.3$, $a_2=12.2$,
$a_3=15.1$, $a_4=18.0$, $a_5=21.0$.

Now, we quantitatively characterize the behavior of geodesics in the
sandwich wave given by (\ref{E9}). The amplitude within the
wavezone is $h=1/\tilde a$ but it follows from Eq. (\ref{E9a})
that we can obtain the corresponding values of $u_s(\phi)$ from
those for $h=1$ (discussed above) by a simple rescaling
$u_s(\phi)=\sqrt{\tilde a}\,u_s^{h=1}(\phi)$, where $u_s^{h=1}(\phi)$
is drawn in Fig.~2. Notice that $u_s\to0$ as $h\to\infty$. It is
obvious that all geodesics for which $u_s(\phi)\le a$, i.e.
$u_s^{h=1}(\phi)\le a/\sqrt{\tilde a}$,  have time enough to
fall into the singularity (in the impulsive
limit, $a=\tilde a\to0$, this condition cannot be satisfied so that
none of the geodesics considered ends in the singularity).
Since $N(\phi)$ is independent of the rescaling of $u$, the
fractal structure of initial conditions given by $I_k$ and the
values of $a_k$ do not depend on $\tilde a$. Consequently, the
condition $ a/\sqrt{\tilde a}\ge a_k$ guarantees that {\it all}
geodesics starting from the initial set $I_k$ will manage to fall into the
singularity at $\rho=\infty$ along the three channels
(bouncing at most $k$-times before that).
For a sequence of sandwich waves with $\tilde a=a$
leading to the impulsive limit as $a\to0$, this condition reduces
to a simple relation $a\ge a_k^2$ which can be interpreted as
follows. In order to emerge the $k$-th level of the fractal
structure of motion the sandwich wave must have at least the duration
$a$ such that $a\ge a_k^2$. For example, if $a>441$ then all
geodesics with $N\le5$ bounces have time enough to choose the
corresponding channel and fall along this into the singularity,
i.e. $I_5$ is fully developed. For smaller $a$ the fractal structure
is `cut' at lower levels $I_k$  so that the geodesic motion becomes
`less chaotic'. For $a<a_0^2\approx40$ the fractal structure vanishes
since there is no time for any geodesic (performing at least one
bounce) to choose the outcome channel and fall into the
singularity. Thus, for $a\to0$ the motion is regular. This
effect occurs in all non-homogeneous sandwich {\it pp\,}-waves with
arbitrary exponent $n=3,4,\cdots$, only the specific values of
$a_k$ for a given $n$ are different: for higher
$n$ the values of $a_k$ are smaller.

\section{Motion in smooth asymptotic sandwich waves}

We have observed that, in non-homogeneous sandwich {\it pp\,}-waves
with a short duration, many particles still remain in the inner
region when the potential (\ref{E7b}) defining the outcome
channels vanishes. These particles move through various points
in different directions having different velocities when the
sandwich wave ends. Subsequently, they move uniformly in the
Minkowski space behind the wave and it is obvious that they will
{\it not} follow former channels to the singularity at $\rho=\infty$.
In fact, any point is accessible by some trajectory.
The narrower the sandwich wave, the greater the number of
particles moving `outside' the channels.

In order to illustrate such behavior of geodesics (and to emphasize
this important aspect of the impulsive limit) we consider another
type of sandwich waves (\ref{E7a}) with the profile function
\begin{equation}
h(u) =\frac{\tilde b}{\cosh^2(2bu)}\ , \label{E10}
\end{equation}
where $b, \tilde b$ are positive constants.
Note that the corresponding radiative spacetimes are
curved everywhere, becoming flat (exponentially fast) only asymptotically as
$u\to\pm\infty$. On the other hand, these waves are smooth since
all derivatives of $h(u)$ are continuous.
Moreover, (\ref{E10}) is more appropriate for numerical integration
of geodesics (contrary to the case of simplest waves given by Eq.
(\ref{E9}) we need not apply the junction conditions which
guarantee the continuity of motion at $u=0$ and $u=a$ where $h(u)$
is discontinuous). Again, we consider a ring of free test particles
which are at rest at $u=0$ (other initial conditions, e.g. a ring
which is at rest at $u=-\infty$ lead to analogous results).
In  Fig.~3 we show their geodesic motion in
the wave with $n=3$ and $h(u)$ of the form
(\ref{E10}), for different values of the parameter $b=\tilde b$.
For very small values of $b$ the profile remains almost constant on
a large interval of $u$ and the trajectories `coincide'
with those shown in Fig.~1. The outcome channels are very well defined
and the motion is chaotic. However, with a growing value of $b$
these channels become fuzzy. It is harder and harder to
distinguish between them, so that the basin boundaries lose their
fractal structure. For $b\to\infty$, the profile function (\ref{E10})
approaches the Dirac delta (in a distributional sense) and again,
in the impulsive limit the geodesic motion is not chaotic but regular.

This effect is quantitatively characterized in Fig.~4 where we
plot the function $j(\phi)$ for  different values of $b$. For
this purpose we define three outcome windows as small
intervals $\Delta\phi$ of angles around the radial axes $\phi_j$
of the three outcome channels (localized sufficiently far away
from the origin); here we consider $|\phi-\phi_j|<\Delta\phi=0.1$.
When a geodesic starting at $\phi$ passes
through some outcome window we assign the corresponding value of
$j(\phi)\in(1, 2, 3)$ to it. If the trajectory does {\it not} pass
through any of the three windows then we define $j(\phi)=0$ which
means that the geodesic does not approach infinity along the
outcome channels $\phi_j$. From a sequence presented in Fig.~4 it
is obvious that the number of geodesics localized outside the
channels increases with a growing value of $b$, i.e. as the
sandwich profiles approach the impulsive limit. In the same
diagrams we also plot the function $u_s(\phi)$. One can observe
that the number of geodesics reaching the singularity in
finite values of $u_s$ rapidly decreases with a growing $b$. In
particular, all geodesics outside the channel windows (with
$j(\phi)=0$) have $u_s(\phi)=\infty$. There is a critical value
$b_c\approx0.09$ such that for $b>b_c$ {\it all} geodesics
considered have $u_s=\infty$, see Fig.~4. These geodesics reaching
$\rho=\infty$ at $u_s=\infty$ are, in fact, given by specific
solutions of Eq. (\ref{E9a}) for which the condition $|h(u)|\ll\rho^{1-n}$
is satisfied for large values of $u$. Consequently,
$|x''|\le|\zeta''|\ll1$,  $|y''|\le|\zeta''|\ll1$, i.e.
the acceleration is negligible and asymptotically the particles move
uniformly along straight lines. The coordinates $x$ and $y$
depend linearly on $u$, so that $\rho=\infty$ is reached at
$u=\infty$. Note that for the smooth profile (\ref{E10}) of the
wave with $n=3$ this condition reduces to
$\cosh(2bu)\gg\sqrt{\tilde b}\,\rho(u)$. Once this condition is
satisfied for some $u$, it is valid for all greater values of $u$
since the left hand side grows exponentially while the right hand
side only linearly as $u\to\infty$.

Finally, we wish to find the critical value of $b_c$.
From Fig.~4 it is obvious that for $b=b_c$
even the geodesic starting at $u=0$ from rest from the unit circle at
$\phi=\pi$ (and similarly from $\phi=\pm \pi/3$) will reach
$\rho=\infty$ with an infinite value of $u_s$. In real coordinates,
this geodesic is given by $x''+h(u)x^2=0$, $y(u)=0$, with $x(0)=-1$
and $x'(0)=0$. Introducing $\psi=\ln|x|$ we get the equation
$\psi''=h(u)\exp(\psi)-{\psi'}^2$, $\psi(0)=0=\psi'(0)$; the
corresponding solutions for various values of $b=\tilde b$ are plotted
in Fig.~5. For small $b$ the function $\psi(u)$ is convex and
diverges at finite values of $u$. For large values of the
parameter $b$, it becomes concave so that the geodesics
approaches infinity only as $u\to\infty$. The boundary between
these two types of behavior defines $b_c$.
From numerical simulations we obtained an approximate value
$b_c=0.0872374$, see Fig.~5.

\section{Motion in impulsive waves}

We have demonstrated above that chaos disappears when the
sequence of non-homogeneous sandwich {\it pp\,}-waves approaches
the impulsive limit. In this section we concentrate on a
description of motion in these impulsive spacetimes. In
particular, we present and discuss an analytic solution to the
geodesic equations, thus proving explicitly that the motion is fully
non-chaotic.

For this purpose it is convenient to use a coordinate system
for impulsive {\it pp\,}-wave spacetimes which is continuous for all values
of $u$ \cite{AB}, \cite{PoVe98}. Following \cite{PoVe98},
the metric can be written in  the form
\begin{equation}
\d s^2=2\,|\d\bar\eta-\textstyle{\frac{1}{2}}u\Theta(u)\,
[d^2F(\eta)/d\eta^2]\, \d\eta|^2-2\,\d u\d r\ .
\label{E11}
\end{equation}
The transformation relating (\ref{E11}) and (\ref{E1}) with
$f=\delta(u)F(\zeta)$ is
\begin{eqnarray}
&&\zeta = \eta-\textstyle{\frac{1}{2}}u\Theta(u)\,(d\bar F/d\bar\eta)
 \ ,\label{E12}\\
&&   v  = r -\textstyle{\frac{1}{2}}\Theta(u)(F+\bar F)
  +\textstyle{\frac{1}{4}}u\Theta(u)\,|d F/d\eta|^2 \ .\label{E13}
\end{eqnarray}
Obviously, there are privileged geodesics
$\eta=\eta_0=$const. in (\ref{E11}) corresponding to free
particles which remain at rest in flat
Minkowski half-space $u<0$ in front of the impulsive wave.
After the passage of the impulse these geodesics are
still given by $\eta=\eta_0$ but the flat half-space $u>0$ behind
the wave is naturally described by coordinates $\zeta$ and $v$
(see (\ref{E1})) in which the motion is uniform.
From Eq. (\ref{E12}) it follows that, for impulsive waves
given by $F(\zeta)=\frac{2}{n}C\zeta^n$ (so that $h(u)=C\delta(u)$),
the motion in the transversal plane is described simply
by $\zeta(u) = \eta_0-C\Theta(u)\, u\,\bar\eta_0^{n-1}$.
If we parametrize the initial position of each particle by
$\zeta(u<0)=\eta_0=\rho_0\exp(\hbox{i}\phi_0)$, the motion behind
the impulse ($u>0$) is explicitly given by
\begin{eqnarray}
&& x(u) = \rho_0\cos\phi_0-C\rho_0^{n-1}\cos[(n-1)\phi_0]\,u \
,\nonumber\\
&& y(u) = \rho_0\sin\phi_0\ +C\rho_0^{n-1}\sin[(n-1)\phi_0]\,u
 \ ,\label{E14}
\end{eqnarray}
so that
\begin{equation}
\rho^2(u)=\rho^2_0+C^2\rho_0^{2n-2}u^2-2C\rho_0^n\cos(n\phi_0)\,u
\ .
\label{E14a}
\end{equation}
The motion is uniform, i.e. the trajectories are straight
lines with the velocity of each particle
$(x',y')=C\rho_0^{n-1}(-\cos[(n-1)\phi_0],\sin[(n-1)\phi_0])$
being constant.
Thus, the inclination of the straight trajectory in the
transversal $(x,y)$-plane is $\alpha=\tan(y'/x')=\pi+(1-n)\phi_0$,
and the speed is $\sqrt{x'^2+y'^2}=|C|\,\rho_0^{n-1}$.
Notice that the speed depends on $\rho_0$ while the direction of
motion on $\phi_0$ only. The above geodesics can easily be
visualized. We do not present their trajectories
which would be very similar to those in sufficiently narrow
sandwich waves, such as the $b=\tilde b=0.5$, $n=3$ case shown in
Fig.~3. Instead, we draw in Fig.~6 `complementary' pictures showing
a deformation of a ring of free test particles (for $\rho_0=1=C$,
$n=3$). The initial circle at $u=0$ is continuously deformed into
smooth curves with $n$ growing loops which are visualized here as
sequences of 4, 8, 13 and 22 consecutive steps $\Delta u=0.1$,
i.e. the largest connected curve describes the deformation of the
ring at $u=2.2$. Similarly, in Fig.~7 the deformation is shown
for $n=3, 4, 5, 6$ and $u=0.25, 0.5, 0.75, 1$.
The most distant particles are those which started at
$\phi_0=\phi_j=(2j-1)\pi/n$, $j=1,2,\cdots,n$ with
$\cos(n\phi_0)=-1$ so that $\rho(u)=|\rho_0+C\rho_0^{n-1}u|$;
they exactly follow the radial axes of the outcome
channels which would be present for corresponding {\it pp\,}-waves
with a constant profile. On the other hand, for $\phi_0$ given by
$\cos(n\phi_0)=1$, we get $\rho(u)=|\rho_0-C\rho_0^{n-1}u|$.

It can be observed that for small values of $u$ the deformation
of the ring agrees with that shown in Fig~8. of Ref. \cite{PVcha2}
describing chaotic motion in the non-homogeneous {\it pp\,}-waves with a
constant profile. The principal difference occurs for large
$u$. In the impulsive case there are no subsequent loops arising
from more and more bounces in the inner region so that the circle can not be
deformed in a fractal way with different segments moving to
different outcome channels. Instead, there are no channels
to the singularity, the trajectory of each particle is explicitly given
by (\ref{E14}), and the motion is non-chaotic.

These results can easily be generalized to the case when the
test particles are not at rest initially. From the metric
(\ref{E1}) it is obvious that, using the coordinate $\zeta$, the
motion must be uniform in both flat half-spaces, i.e.
$\zeta(u<0)=\psi_0 u+\eta_0$, $\zeta(u>0)=\chi_0 u+\zeta_0$,
where $\psi_0, \eta_0, \chi_0, \zeta_0$ are constants. The
relation between these parameters can again be found using the
continuous form of the impulsive metric (\ref{E11}). By solving
the corresponding geodesic equations we could obtain $\eta(u)$.
Although this function is very complicated for $u>0$, it  has the
property that $\eta(u)$ and $\eta'(u)$ are continuous (even at
$u=0$). Using this fact and Eq. (\ref{E12}), which for
$F(\zeta)=\frac{2}{n}C\zeta^n$ reduces to
$\zeta(u>0) = \eta(u)-C\, u\,\bar\eta^{n-1}(u)$, we obtain
$\lim_{u\to0}\zeta(u>0)=\zeta_0=\eta(0)=\eta_0$ and
$\zeta'(u>0)=\chi_0=\eta'(0)-C\,\bar\eta^{n-1}(0)
=\psi_0-C\,\bar\eta^{n-1}_0$, i.e.
\begin{equation}
\zeta(u)=[\psi_0-C\Theta(u)\bar\eta^{n-1}_0]\,u+\eta_0\ .
\label{E15}
\end{equation}
Therefore, the trajectory of each particle in the transversal plane
is a continuous but refracted straight line with a discontinuity
in the velocity at $u=0$ given by
$\Delta\zeta'\equiv\chi_0-\psi_0=-C\,\bar\eta^{n-1}_0$.
Notice that the value of the jump depends on the position $\eta_0$
of the particle at $u=0$ only, not on its actual velocity.
By parametrizing $\eta_0=\rho_0\exp(\hbox{i}\phi_0)=\zeta_0$ we
immediately obtain the discontinuity in velocity,
$(\Delta x',\Delta y')=C\rho_0^{n-1}(-\cos[(n-1)\phi_0],\sin[(n-1)\phi_0])$.
Thus, the result is trivial: if the particle is
not at rest initially, behind the wave its constant velocity merely
superimposes to the effect of a characteristic jump in velocity given by the
impulse. By denoting
$\cot\alpha_x\equiv x'(u<0)={\cal R}e\,\psi_0$,
$\cot\beta_x\equiv -x'(u>0)=-{\cal R}e\,\chi_0$, and similarly
$\cot\alpha_y\equiv y'(u<0)= {\cal I}m\,\psi_0$,
$\cot\beta_y\equiv -y'(u>0)=-{\cal I}m\,\chi_0$, we can rewrite
the expression for $\Delta\zeta'$ as
$\cot\alpha_x+\cot\beta_x=C\rho_0^{n-1}\cos[(n-1)\phi_0]$ and
$\cot\alpha_y+\cot\beta_y=-C\rho_0^{n-1}\sin[(n-1)\phi_0])$,
which generalize to non-axisymmetric cases the `refraction
formula' for deflection on (null) geodesics  in the axisymmetric
Aichelburg \& Sexl spacetime \cite{AiSe} (see e.g. \cite{DH},
\cite{FPV}, \cite{HS}).

So far, we have concentrated on a description of motion in the transversal
plane. We should also comment on behavior in the
longitudinal direction. From the form of the metric (\ref{E11}) it
follows that, for geodesics $\eta=\eta_0$, the
coordinate $r(u)$ is given by $r(u)=s_0 u+r_0$, $s_0, r_0$ being
constants. Moreover, $r(u)$ is continuous for all $u$.
Due to the Eq. (\ref{E13}), there is a
jump in
\begin{equation}
v(u)=s_0 u+r_0+C^2\rho_0^{2n-2}u\Theta(u)
-\textstyle{\frac{2}{n}}C\rho_0^n\cos(n\phi_0)\,\Theta(u)
\label{E16}
\end{equation}
at $u=0$ given by $\Delta v=-\frac{2}{n}C\rho_0^n\cos(n\phi_0)$
depending both on $\rho_0$ and $\phi_0$. Since the Minkowski
coordinates  are $z_M(u)=\frac{1}{\sqrt2}[v(u)-u]$,
$t_M(u)=\frac{1}{\sqrt2}[v(u)+u]$, there is a discontinuity
$\Delta z_M=\Delta t_M=\frac{1}{\sqrt2}\Delta v$ at $u=0$.
This `shift' effect of impulsive wave on geodesics is well-known
for the axisymmetric Aichelburg \& Sexl spacetime \cite{AiSe},
see \cite{DE78}, \cite{DH}, \cite{FPV}, \cite{HS} and elsewhere.

A solution of the geodesic (and geodesic deviation) equations in general
impulsive {\it pp\,}-wave spacetime has been presented
in \cite{Bal97}-\cite{KuSt98}. Starting from the distributional
form of the metric (\ref{E1}), one has to deal with
ill-defined products of distributions. In order to obtain correct
results in a mathematically rigorous fashion, careful
regularization procedures and delicate manipulations during
calculation of the distributional limit are required. In fact,
the Colombeau theory of generalized functions providing a
suitable consistent framework has to be applied. Using this
rigorous solution concept, it was shown in \cite{KuSt98} that in
the impulsive limit the geodesics are totally independent of the
regularization, i.e. on the particular shape of the sandwich wave
(the impulsive wave `totally forgets its seed').
Here we obtained an identical explicit form of geodesics starting from
the continuous form of the impulsive metric (\ref{E11}).
It was demonstrated in \cite{KuSt99} that these two approaches
are equivalent in a mathematical sense (even if the transformation
(\ref{E12}), (\ref{E13}) is discontinuous). However, our
main goal here was not to re-derive the known geodesics but
to concentrate on their physical description and visualization.

\section{Focusing of geodesics and caustic properties}

Obviously, geodesics with parallel trajectories in flat
half-space in front of a sandwich or impulse wave are refracted.
A natural question arises whether a specific character of the refraction
leads to some form of focusing of the corresponding geodesics. Such
an astigmatic focusing effect is well-known for plane waves,
and results in interesting
caustic properties thoroughly investigated in \cite{BP}.
For principal reasons (non-existence of a diagonal Rosen form of
the metric, see Eq. (15) in \cite{KuSt99}, or the presence of chaos in
geodesic motion) it would be a very difficult task to reproduce
these results for non-homogeneous sandwich waves. Therefore,
we restrict ourselves to impulsive gravitational waves only.

It immediately follows from Eqs. (\ref{E14}), (\ref{E16})
that particles staying at fixed $x_0, y_0, z_0$ in front of the
impulse, move at $u>0$ according to
\begin{eqnarray}
&&x_M(u)=\sqrt2\,x(u)\ ,\qquad y_M(u)=\sqrt2\,y(u)\ ,\label{E17}\\
&&\sqrt2\,[z_M(u)-z_0]=C^2\rho_0^{2n-2}u
-\textstyle{\frac{2}{n}}C\rho_0^n\cos(n\phi_0)\ ,\label{E18}
\end{eqnarray}
where $x_M, y_M, z_M$ are Minkowski coordinates
behind the impulse, $x(u)$ and $y(u)$ are given by (\ref{E14}).
In particular, for impulsive plane waves ($n=2$) we get
$x_M(u)=(1-Cu)x_0$, $y_M(u)=(1+Cu)y_0$ and
$2\sqrt2\,[z_M(u)-z_0]=C(1+Cu)y_0^2-C(1-Cu)x_0^2$, where
$x_0=\sqrt2\rho_0\cos\phi_0$, $y_0=\sqrt2\rho_0\sin\phi_0$.
Consequently, at $u=u_f\equiv1/C>0$ one gets $x_M(u_f)=0$, $y_M(u_f)=2y_0$
and $\sqrt2\,[z_M(u_f)-z_0]=Cy_0^2$, so that all particles meet
at $x_M=0$, independently of their initial position
$x_0$ in the $z_0$-plane. The generically quadratic surface
(hyperbolic paraboloid) $u=$const. degenerate at $u=u_f$ to a parabolic
caustic line $x_M=0$, $z_M=\frac{C}{4\sqrt2}y_M^2+z_0$. This is shown in
Fig.~8 which visualizes the coordinate singularity formation.
It plays a crucial role in colliding  plane-wave solutions
where  the corresponding global `fold' singularities arise, as described
in \cite{MT}. Consequently, these spacetimes are globally
hyperbolic in contrast to single plane-wave solutions
where the focusing effect on null cones forbids an existence of a
spacelike hypersurface which would be adequate for the global
specification of Cauchy data  \cite{Penr}.

Here we wish, however, to investigate key features of the focusing effect
in more general non-homogeneous impulsive {\it pp\,}-wave spacetimes
with $n=3,4,5,\cdots$.
From the deformation of a ring of free particles indicated in Fig.~7
it follows that {\it only $n$ privileged particles can collide} at
the origin (in contrast to the plane-wave case where particles with
arbitrary $x_0$ are focused along the $x_M=0$ line). It can easily be
derived from Eq. (\ref{E14a}) by setting $\rho(u)=0$ that only those
$n$ particles with $\cos(n\phi_0)=1$  reach the origin of the
transversal ($x,y$)-plane, simultaneously at $u=u_f\equiv\rho_0^{2-n}/C$.
Notice that particles starting from a larger circle focus sooner. This
is understandable: they have to travel a bigger distance $\rho_0$, however,
their speed is also bigger, $C\rho_0^{n-1}$, so that the travel time is
indeed $\rho_0^{2-n}/C$. Note that the occurence of such focusing is
associated with the coordinate singularity in the Rosen form of
the metric (\ref{E11}).

At $u=u_c$ the circle is deformed into a curve with $n$ cusps, see Fig.~7.
The value of $u_c$ can be found from the condition
$(\partial\phi/\partial\phi_0)|_{\cos(n\phi_0)=-1}=0$, where
$\phi\equiv\arctan [y(u)/x(u)]$, with $x(u), y(u)$ being given by
(\ref{E14}). This leads to equation
$(n-1)C^2\rho_0^{2n-2}u_c^2+(n-2)C\rho_0^nu_c-\rho_0^2=0$
admitting a unique positive solution $u_c=\rho_0^{2-n}/[C(n-1)]
=u_f/(n-1)$. For $\rho_0=1=C$ we get a simple expression $u_c=1/(n-1)$
which is in agreement with Fig.~7.

In fact, the shape of deformation of the initial circle $\rho=\rho_0$
of test particles in the transversal plane has a beautiful
geometrical interpretation: at any time $u$, {\it the circle is
deformed into a curve called a hypotrochoid}. Such a curve is
generated by a point $p$ attached to a small circle of radius $B$
rolling around the inside of a large fixed circle of radius $A$,
with $H$ being the distance from $p$ to the centre of the rolling
circle \cite{Weis}. The Eqs. (\ref{E14}) for particle motion
behind the impulsive wave are just the parametric equations for a
hypotrochoid with the identification
\begin{equation}
A=\frac{n}{n-1}\,\rho_0\ ,\qquad B=\frac{1}{n-1}\,\rho_0\ ,\qquad
H=Cu\rho_0^{n-1}\ ,
\label{E20}
\end{equation}
where $\phi_0$ is the parameter. We observe that $A=nB$ so that
 {\it the hypotrochoids are closed curves with $n$ loops} in our case.
Also, the parameter $H$ grows linearly with $u$. At $u=0$
we have $H=0$ and the curve is just the initial circle of
radius $A-B=\rho_0$. At $u=u_c$ the parameter has the value
$H=\rho_0/(n-1)=B$ and the hypotrochoid reduces to the curve
called a {\it hypocycloid} which has $n$ cusps. Note that
it degenerates for $n=2$ to a line segment, a 3-cusped hypocycloid  is
called a deltoid, a 4-cusped hypocycloid is called an
astroid (see Fig.~7 for $n=3$ and $n=5$ case). Finally, at
the focusing time $u=u_f>u_c$ we get $H=\rho_0=A-B$ so that the
hypotrochoid $n$-times intersects the origin. The corresponding
curve is called a {\it rose} since it resembles a flower with $n$
petals (for $n=2$  the rose degenerates to a line since $u_f=u_c$).

It should be emphasized that the above behavior of geodesics
in the  $(x,y)$-plane describes only part of the
overall deformation. We have to consider not only the transversal
deformation but also the motion in the
$z_M$-direction described by (\ref{E18}), as we have done for the
plane-wave case in Fig.~8. At $u=0$ the longitudinal deformation
$\sqrt2\,[z_M(u)-z_0]$ caused by the impulse is given by the second
term on the right hand side of Eq. (\ref{E18}), which is exactly the
shift $\Delta v=-\frac{2}{n}C\rho_0^n\cos(n\phi_0)$. For $u>0$
the first term increases with $u$; at
$u_w=\frac{2}{n}\rho_0^{2-n}/C=\frac{2}{n}u_f$ the deformation of the
whole disc $\rho\le\rho_0$, $z_0=$const., becomes non-negative and subsequently
grows to positive values, uniformly with $u$. In combination with
the deformation in the transversal plane we get an interesting new effect
of `wrapping up' of the inner part of the disc into a closed caustic surface
as shown in Fig.~9. Note that this behavior is specific for
non-homogeneous {\it pp\,}-waves and does not occur in impulsive
plane waves since $u_w=u_f$ when $n=2$.

\section{Conclusions}

We have shown for non-homogeneous {\it pp\,}-wave
spacetimes that the motion of free test particles is chaotic.
However, for non-homogeneous sandwich gravitational waves the chaotic
behavior smears as the duration of the wave tends to zero. In the
limit of impulsive waves the motion is regular. The focusing
effect of non-homogeneous impulsive waves can also be described
explicitly and leads to a formation of a specific closed caustic
surface. It would be an interesting
task to investigate from this point of view geodesics in other
classes of exact radiative spacetimes and elucidate an open
question whether such a type of behavior is common in general relativity.

Also, it is obvious that sandwich {\it pp\,}-wave solutions may serve as
local models of gravitational waves far away from a radiating
source. In realistic situations these waves should contain
non-homogeneous components whose influence on test particles has
been described above. In particular, chaotic-type effects on null
geodesics could in principle be observable after the passage of the
sandwich wave as characteristic ``chaotic deformation patterns'' of an
observer's view of the sky, i.e. as peculiar changes of positions
of stars and galaxies. These specific astronomical consequences
will be investigated elsewhere.

\section*{Acknowledgments}

We acknowledge the support of grant no GACR-202/99/0261
from the Czech Republic.
We thank the developers of the software system FAMULUS
(Tom\'a\v s~Ledvinka in particular) which we used for
computation and drawing of all the pictures. Also, we thank Jerry
Griffiths for all his help with the manuscript.

\newpage

{\LARGE Figure Captions:}
\bigskip

{\bf Fig.1}
Geodesics starting from a unit circle escape to infinity
only along one of the $n$ channels (left).
The functions $j(\phi)$ and $\tau_s(\phi)$
indicate that basin boundaries separating different outcomes
are fractal (right).
\smallskip

{\bf Fig.2}
Plots of the functions $N(\phi)$ and $u_s(\phi)$ in the
sequence of ever narrower intervals exhibit their fractal
structure. Obviously, the number $N$ of bounces of each geodesic
is an appropriate measure of the fractal level.
\smallskip

{\bf Fig.3}
Geodesics in the smooth sandwich gravitational
waves given by  Eq. (10), $n=3$. For small values of the
parameter $b=\tilde b$, most of them escape to infinity
only along the well-defined three channels.
For larger values of $b$ these channels become fuzzy
and the basin boundaries separating different outcomes
are losing their fractal structure. In the impulsive limit
($b\to\infty$) the motion is regular.
\smallskip

{\bf Fig.4}
Plots of the functions $j(\phi)$ and $u_s(\phi)$
describing behavior of geodesics in smooth sandwich gravitational
waves. When the parameter $b=\tilde b$ is small, most geodesics reach the
singularity in finite value of $u_s$. For $b>b_c\approx0.0872374$
all geodesics have $u_s=\infty$. Also, for higher $b$ the number
of geodesics outside the three outcome windows ($j=0$) is greater.
\smallskip

{\bf Fig.5}
The function $\psi(u)=\ln|x(u)|$ describing specific
geodesics changes its character at the critical value
$b_c\approx0.0872374$. For $b>b_c$ the geodesics
approaches $\rho=\infty$ only as $u\to\infty$.
\smallskip

{\bf Fig.6}
The deformation of a ring of free test particles which
are at rest initially in the tranvsveral $(x,y)$-plane
by the influence of an impulsive gravitational wave with
$f=\frac{2}{3}\delta(u)\zeta^3$ at $u=0.4, 0.8, 1.3$ and $2.2$ with
steps $\Delta u=0.1$. Bellow is the 3D visualization of the
deformation $(x(u), y(u))$.
\smallskip

{\bf Fig.7}
Deformation of a unit ring of particles by impulsive
waves with $f=\frac{2}{n}\delta(u)\zeta^n$ at $u=0.25, 0.5, 0.75, 1$
for $n=3, 4, 5, 6$. At $u=u_f\equiv\rho_0^{2-n}/C=1$ all the privileged
$n$ particles focus at the origin of the transversal plane. The
curves are hypotrochoids, as explained further in the text.
\smallskip

{\bf Fig.8}
Deformation of a disc consisting of free particles
$\rho_0\le1$, $z_0=0$. By an impulsive plane wave $(n=2, C=1)$
the dics is deformed into a hyperbolic paraboloidal surface
at $u=$const.$\ge0$ in the Minkowski space behind the wave. At
$u=u_f=1$ the surface degenerates to a caustic parabolic line.
\smallskip

{\bf Fig.9}
Deformation of a disc $\rho_0\le1$, $z_0=0$ of particles
by an impulsive non-homogeneous wave $f=\frac{2}{3}\delta(u)\zeta^3$
at $u=0, 0.25, 0.5, 0.75, 1$ and $1.25$ in the Minkowski space behind
the wave. At $u=u_f=1$ all the privileged three
particles collide and with the deformation in the longitudinal
$z_M$-direction a closed caustic surface is created.

\end{document}